\documentclass{ifacconf}

\usepackage{graphicx}      
\usepackage{natbib}       
\usepackage{amsmath}
\usepackage{breqn}
\usepackage{xcolor}
\usepackage{physics}
\usepackage{mathrsfs}
\usepackage{amsfonts}
\usepackage{algorithmic}
\usepackage{textcomp}
\begin{document}
\begin{frontmatter}

\title{A Fast Diagnostic to Inform Screening of Discarded or Retired Batteries } 

\author[First]{Joseph A. Drallmeier} 
\author[First]{Clement Wong} 
\author[First]{Charles E. Solbrig}
\author[First]{Jason B. Siegel}
\author[First]{Anna G. Stefanopoulou}

\address[First]{Battery Controls Group, University of Michigan, Ann Arbor, MI, 48109 USA. (e-mail: drallmei, clemwong, csolbrig, siegeljb, annastef@umich.edu).}

\begin{abstract}                
With the increased pervasiveness of Lithium-ion batteries, there is growing concern for the amount of retired batteries that will be entering the waste stream. Although these batteries no longer meet the demands of their first application, many still have a significant portion of their initial capacity remaining for use in secondary applications. Yet, direct repurposing is generally not possible and each cell in a battery must be evaluated, increasing the cost of the repurposed packs due to the time intensive screening process. In this paper, a rapid assessment of the internal resistance of a cell is proposed. First, this method of measuring the resistance is completed on cells from twelve retired battery packs and one fresh pack using a hybrid pulse power characterization (HPPC) test as a benchmark for the analysis. Results from these tests show relatively constant resistance measurements across mid to high terminal voltages, allowing this metric to be independent of state of charge (SOC). Then, the relation between internal resistance and capacity across the various packs is discussed. Initial experimental results from this study show a correlation between internal resistance and capacity which can be approximated with a linear fit, suggesting internal resistance measurements taken above a threshold cell terminal voltage may be a suitable initial screening metric for the capacity of retired cells without knowledge of the SOC.
\end{abstract}


\end{frontmatter}

\section{Introduction}

The current exponential increase in the production of lithium-ion batteries for clean energy and electric vehicle applications will inevitably result in the presence of a large volume of “retired” batteries soon. Estimates from \cite{IEA2020} suggest 100-120 GWh of EV batteries will be retired by 2030. Batteries are retired when they are unable to meet the original application's performance requirements. A battery pack for an electric vehicle, for instance, is under warrantied to maintain  80\% of its original capacity (\cite{Groenewald2017}, \cite{Cusenza2019}) within ten years or 100,000 miles travelled with the battery. Retired batteries, while no longer able to meet performance standards of the the original application, still retain much of their performance capabilities in most cases and are sufficient to be applied in less-demanding applications (\cite{Zhu2021}). As such, to maximize the environmental and economic benefits of lithium-ion batteries, remanufacturing and repurposing is a key step in a battery's life-cycle (\cite{Hua2020}).

However, the cost of testing and refurbishing batteries is high and may prevent applications of retired batteries for second life from being economically viable (\cite{Rallo2020}, \cite{MartinezLaserna2018}). \cite{Neubauer2015} indicates the greatest portion of the testing and refurbishing cost is the laborer costs, much of which is due to the time spent disassembling and testing a battery. A significant contributor to long test times is the assessment of a battery's state-of-health (SOH), which is a measure of a battery's current performance conditions compared to its original state. For instance, assessing battery capacity fade, a key metric in SOH, is conventionally completed by evaluating changes in the open circuit voltage (OCV) with respect to state of charge (SOC) (\cite{Roscher2011}). However, obtaining the OCV curves takes on the order of $20+$ hours.
Further, direct testing and repurposing batteries at the pack level typically depends on the active balancing and more time intensive measures of evaluating individual cells is required. As noted by \cite{Zhou2017}, variations in individual cells making up a battery pack increase with aging of the pack and a few individual cells can reduce the performance of the entire pack. Thus, developing a strategy that reduces testing time while providing an acceptable estimate of a cell's capacity could significantly reduce cost of testing batteries and make applications of retired batteries more economically feasible.

In (\cite{Zhou2020}), the current methods for evaluating retired or aged battery packs are segregated into two methodologies. The first focuses on the effectiveness and consistency of the evaluation process while neglecting time and effort considerations. An example of this is presented by \cite{Chung2021} in which the standard method of evaluating the safety and performance of a retired battery is completed, as outlined by the Underwriters Laboratories (UL), a global safety certification company, in UL 1974. The second methodology, which is where this study aims to contribute, focuses on a surrogate the OCV that can be aquired faster than the current state of the art of $20+$ hours while still maintaining an acceptable level of consistency. \cite{Zhou2020} used incremental capacity curves obtained with high charging current rates which are then used to extract approximations of capacity and internal resistance of cells. However, a full charge of the cells is still required despite occurring at a faster rate. \cite{Weng2013} proposes a battery's capacity estimation scheme that only needs charging data from 60\%–85\% SOC to estimate the battery capacity and avoids the need for a time-consuming full charge and discharge. Recent work by \cite{Mohtat2022} shows that this range of SOC provides critical information that correlates well with a cell's SOH, yet a full charge for one cell is typically required to reset SOC evaluation when the initial SOC is unknown.

Advancing this state of the art, the following study investigates relation between an estimated internal resistance parameter, $R_s$ and the measured capacity of retired batteries. While internal resistance is only one of many parameters to influence the performance of a battery (\cite{Zhou2017}), it may be a rapid metric with which to initially screen retired batteries. In the following sections, the testing procedures used to find the cell capacity as well as to rapidly estimate internal resistance through charge interrupt are provided. Additionally, a hybrid pulse power characterization (HPPC) test is completed to substantiate the internal resistance estimates. Then, these tests were performed on discarded DeWALT lithium-ion power tool battery packs and a discussion of the $R_s$ characteristics is provided. From these results, a clear correlation between cell capacity and resistance is observed. Further, these results hold across mid to high range terminal voltage values. This evaluation technique can then be used once surpassing a threshold cell terminal voltage without any knowledge of SOC, providing a fast initial screening test for aged Lithium-ion batteries.

\section{Battery Packs}
\begin{figure}[t]
\begin{center}
\includegraphics[width=8.8cm]{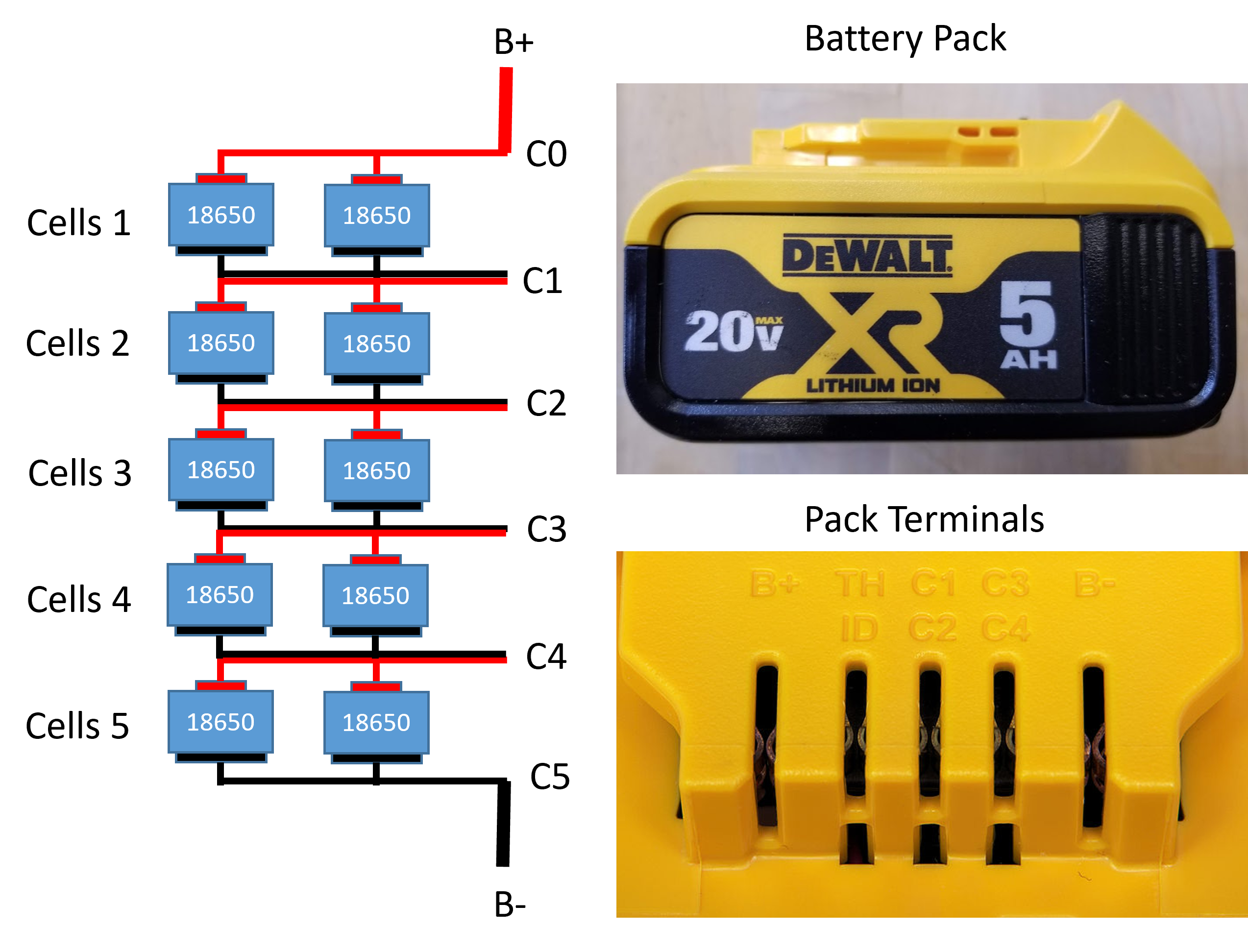}    
\caption{Schematic of the 2P5S test setup along with an image of the battery pack and pack terminals.} 
\label{fig:Packs}
\end{center}
\end{figure}
The battery packs used in this study were DeWalt 20 V, 5 A-h model DCB205 lithium ion power tool battery packs, with a total of thirteen battery packs analyzed. An image of the pack as well as a schematic of the 2P5S cell arrangement within the pack is provided in Fig. \ref{fig:Packs}. Each cell is an 18650 cylindrical cell with a capacity of 2.5Ah. It should be noted here that throughout the rest of the paper, references made to cells 1 through 5 denote the cell pairs in parallel, as shown in Fig. \ref{fig:Packs}. 

Twelve of the packs are considered aged and were acquired from a recycling center after consumers discarded them, presumably when deemed past their usable lifespan. The use and cycling history of these batteries are unknown. A fresh pack was purchased to act as a control test. The aged packs were grouped according to the two primary ID codes found, being N330105, which was determined to contain Samsung INR18650-25R cells with an NCA chemistry, and N437615 which contained LG LGABHE21865 cells with an NMC chemistry. Additionally, the new pack had an ID code of N522573 which also contains cells manufactured by LG with an NMC chemistry. In the following discussions, letters are used to denote the various packs. Pack A denotes the fresh pack while all other letters are the retired packs.

\section{Experimental Setup}
The tests sequences and data acquisition were completed with an A\&D iTest Test Cell System. The iTest System has an analog input module used for acquisition of the individual cell pair data voltage. A Bitrode FTV-1 Power Module with 3 current ranges (2A,20A,200A) controlled by the iTest system was the source of the voltage and current profiles. 

The following sections outline the main testing procedures for the battery packs followed by an additional HPPC sequence. The estimated resistance parameters from the discharge interrupt, charge interrupt, and HPPC are compared to demonstrate the validity of the more rapid diagnostic procedure of interrupting a charge or discharge without providing that battery an extended rest period. The parameter of interest here is the ohmic resistance, $R_s$, which is responsible for the instantaneous voltage change when the cell is under load.


\subsection{Testing Procedure} \label{sec:test_procedure}
\begin{figure*}[t]
\begin{center}
\includegraphics[width=18cm]{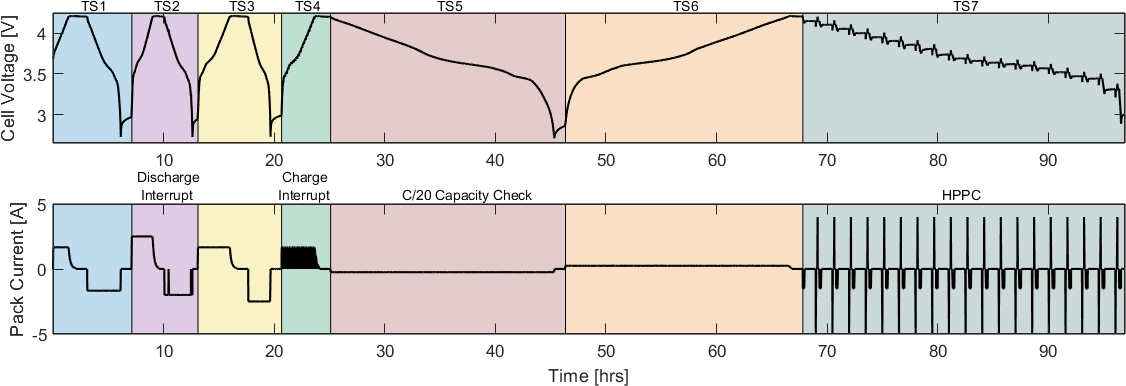}    
\caption{Cell Voltage and pack current profiles used for full characterization of each battery pack. Test sequences (TS) 1-6 are described in Tab. \ref{tb:Test_parameters} and the discharge interrupt of TS2, charge interrupt of TS4, and HPPC of TS7  are detailed in section \ref{sec:test_procedure}.}
\label{fig:test_profile}
\end{center}
\end{figure*}
The cell voltage and pack current profile initially used for evaluation of each battery pack is shown in Fig. \ref{fig:test_profile}. The profile can be separated into seven distinct test sequences (TS), labeled above the figure, with a 30 minute rest period between each individual test. The parameters for each sequence are listed in Tab. \ref{tb:Test_parameters}. The charge and discharge rate for each sequence are listed as C-rates, with the units of $h^{-1}$ and defined as $C \equiv \frac{I}{Q_{p}}$ where $I$ is the current required to charge or discharge the pack ampere hour capacity $Q_p$ in the specified time. In TS1, an initial charge and discharge is completed for initial cycling of the pack as the initial state of charge (SOC) is unknown. The TS2 contains two current interrupts, which will be reffered to as discharge interrupts, occuring at 4V and 3.2V, to provide a means of calculating $R_{s,DI}$ from the instantaneous voltage change as
\begin{equation} \label{eq:R_S}
R_s = \frac{\Delta V_s}{\Delta I}
\end{equation}
where $V_s$ is considered to be the instantaneous voltage change when current is interrupted.
During these discharge interrupts, the sampling rate is 1 Hz. The following TS3 simply provides an additional charge and discharge to isolate TS2 from TS4. In TS4, a charge interrupt procedure is completed across a range of voltages with a sampling rate of 10 Hz. A charging rate of $C/3$ is interrupted at 560 second intervals during which the current is set to 0 for 20 seconds. This allows $R_{s,CI}$ to be calculated over a range of cell voltages again using (\ref{eq:R_S}).

\begin{table}
\begin{center}
\caption{Key details for each test sequence.}\label{tb:Test_parameters}
\begin{tabular}{ccc}
 & Charge Rate & Discharge Rate \\\hline
TS1        & C/3 & C/3    \\
TS2        & C/2 & C/2.5 \\ 
TS3        & C/3 & C/2   \\ 
TS4        & C/3 w/ interrupt & N.A. \\ 
TS5        & N.A & C/20   \\ 
TS6        & C/20& N.A. \\ 
TS7        & N.A.& C/3.33 w/ interrupt \\
\hline
\end{tabular}
\end{center}
\end{table}

The purpose of TS5 is to find the capacity of each cell. Ideally, each charging profile is meant to follow a constant current, constant voltage (CCCV) charging strategy. From this fully charged initialization point in TS5, coulomb counting could be utilized to find cell capacity during the slow discharge of TS5. This slow discharge reduces the impact of internal resistance and concentration gradients on the measured terminal voltage, allowing for an accurate estimate of capacity by assuming the measured terminal voltage approximates the open circuit voltage (OCV) of the battery. However, due to imbalances in the individual cell voltages as the battery pack relies on an external balancing circuit, constant voltage charging could not always be achieved. Therefore, capacity is still measured using coulomb counting, but only between voltage set points of 4.1V and 2.95V.

\begin{figure}[t]
\begin{center}
\includegraphics[width=8.4cm]{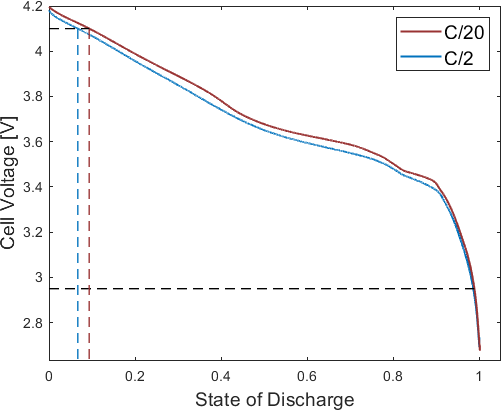}    
\caption{Comparison of measured cell terminal voltage during different discharge rates. When estimating capacity between two voltage points, the voltage drop due to resistance can heavily impact estimates with high discharge rates, increasing the estimated capacity.} 
\label{fig:Cap_measure}
\end{center}
\end{figure}

Using this method of quantifying capacity between voltage limits introduces a non-intuitive consequence of increasing the measured capacity for higher discharge rates. As discharge rate increases, measured capacity becomes more dependent on Rs due to the initial drop in terminal voltage, as shown in Fig. \ref{fig:Cap_measure}. However, by using the slow discharge rate, this relation is small and can be neglected. Therefore, a slow discharge rate is imperative for a reliable estimation of capacity, especially when using this method of measuring between voltage limits.

\begin{figure}
\begin{center}
\includegraphics[width=8.85cm]{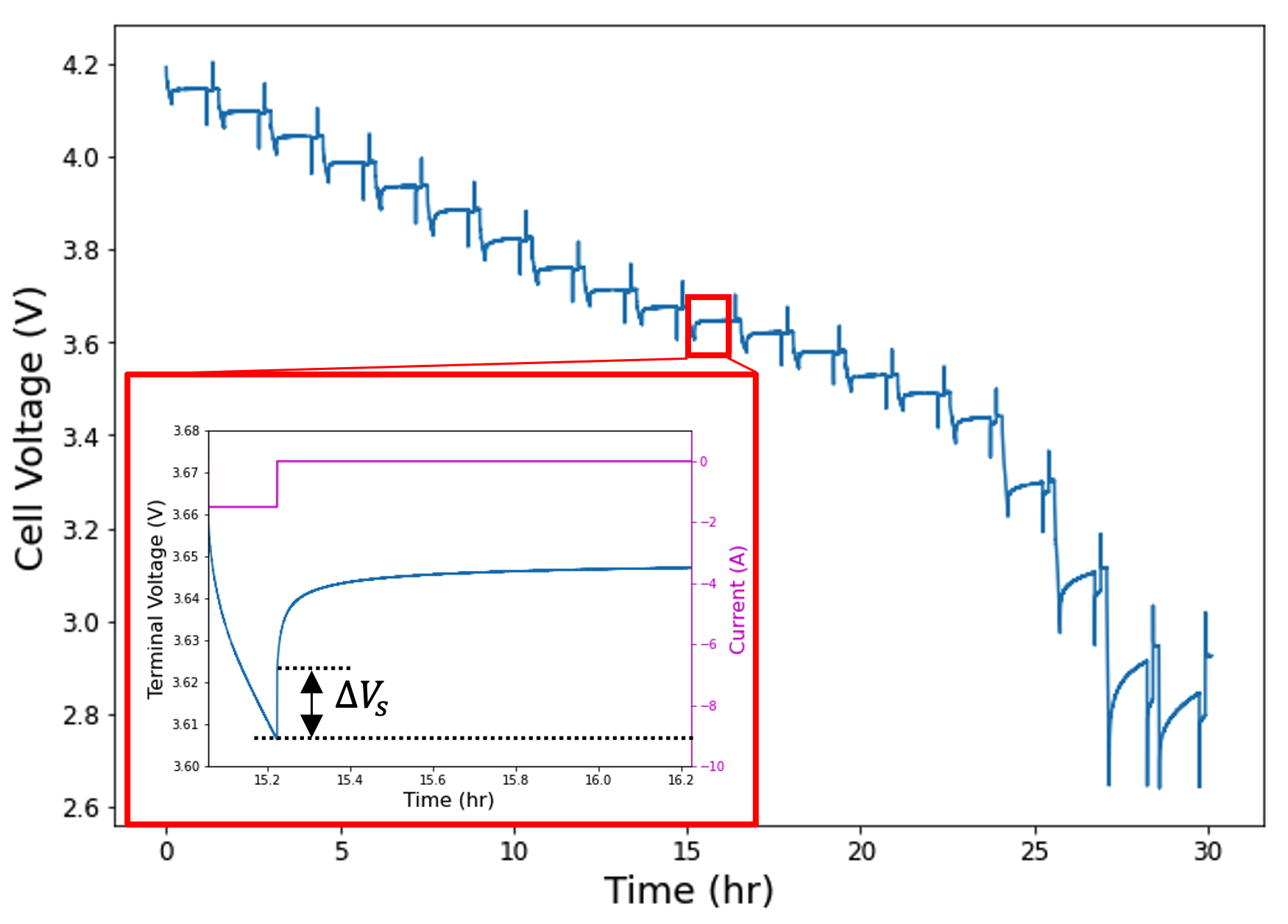}    
\caption{Profile of the cell voltage during the HPPC test sequence. During each interupt of the discharge current, the instantaneous as well as total voltage change of the cell is recorded to calculate the internal resistance parameters.} 
\label{fig:HPPC}
\end{center}
\end{figure}

All packs also underwent an HPPC testing sequence, TS7, where each cell was subjected to series of 10-minute C/3.33 constant-current discharge pulses, followed by a 60-minute rest period. This test was chosen as it provides a benchmark industry standard test procedure for characterizing the batteries internal resistance parameters (\cite{Zhang2011}). 
Figure \ref{fig:HPPC} provides a cell’s voltage response to the current pulses of the HPPC testing sequence which was recorded at a rate of 10 Hz. At every discharge pulse, $R_{s,HPPC}$ was calculated using (\ref{eq:R_S}). The response of the voltage to a step change in current during the HPPC test is shown in Fig. \ref{fig:HPPC}, with the change in voltage due to the ohmic resistance denoted as $\Delta V_s$.

The methods utilized to calculate $R_{s,DI}$, $R_{s,CI}$, and $R_{s,HPPC}$ from a step change in current are identical. Yet, should be noted that while these methods attempt to capture the instantaneous voltage change, it is unavoidable that time will elapse between the discrete sampling points, therefore capturing some of the more rapid diffusion dynamics in the $R_s$ value. As the sampling times when finding $R_{s,CI}$ and $R_{s,HPPC}$ are identical, a direct comparison of these results is valid. However, the larger sample time used when calculating $R_{s,DI}$ increases the resistance value.

\section{Results and Discussion}
The following sections present the $R_s$ parameters calculated from the introduced test sequences and outline the possible efficacy of  the $R_s$ parameter in the estimation of the remaining capacity of a lithium-ion battery. The characteristics of $R_s$ over a range of terminal voltages is discussed along with a comparison of the $R_s$ parameters calculated by the charge interrupt and HPPC test sequences as well as a brief discussion of the discharge interupt results to convey the importance of a standardized sampling time when implementing this method. Additionally, the apparent linear relation between $R_s$ and battery capacity, a key factor of the SOH the battery, is introduced despite the relatively small sample set of aged batteries in this study.

\subsection{Series Resistance Parameters}

\begin{figure}
\begin{center}
\includegraphics[width=8.4cm]{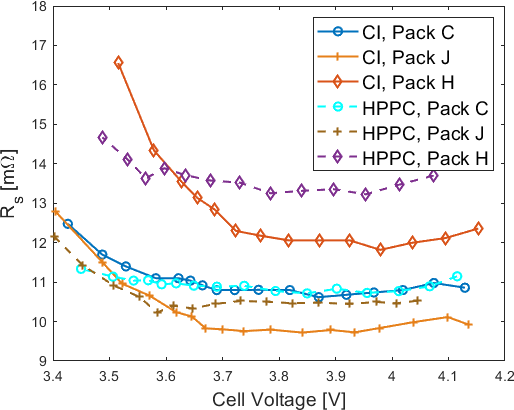}    
\caption{Results of the $R_s$ calculations for the charge interrupt (CI) and HPPC test sequences over a range of terminal voltages for cell 4 in battery packs C, J, and H. } 
\label{fig:HPPCvCI}
\end{center}
\end{figure}

As the HPPC test sequence provides a benchmark to the industry standard practice for parameterizing internal resistance in batteries, a comparison between the $R_{s,HPPC}$ calculated from the HPPC test and the $R_{s,CI}$ calculated from the charge interrupt is provided in Fig. \ref{fig:HPPCvCI} for cell 4 in three different aged battery packs. The packs C and J were chosen as pack C contains cells manufactured by LG, while pack J contains cells manufactured by Samsung. In addition, pack H was included as this pack showed much more significant signs of aging as compared to the rest of the packs studies, as will be discussed later. 

From Fig. \ref{fig:HPPCvCI} it is apparent that while there is not perfect alignment of the results from the charge interrupt and HPPC tests, the trend in $R_s$ between the different packs is maintained. The resistance calculated in the cell from pack H is considerably larger than that of packs C and J. It should also be noted that while each testing sequence calculates $R_s$ in a similar manner, the charge interrupt was completed during charging of the battery while the HPPC test was completed during a discharge, which may account for some of the difference in $R_s$ values. Furthermore, as the HPPC test attempts to characterize not only the ohmic resistance $R_s$ but also the dynamic resistance, an extended rest period is required, resulting in a test lasting approximately 30 hours in this case. The charge interrupt test lasted only 3 hours, an order of magnitude reduction in time required to complete the test. 
\begin{figure}
\begin{center}
\includegraphics[width=8.4cm]{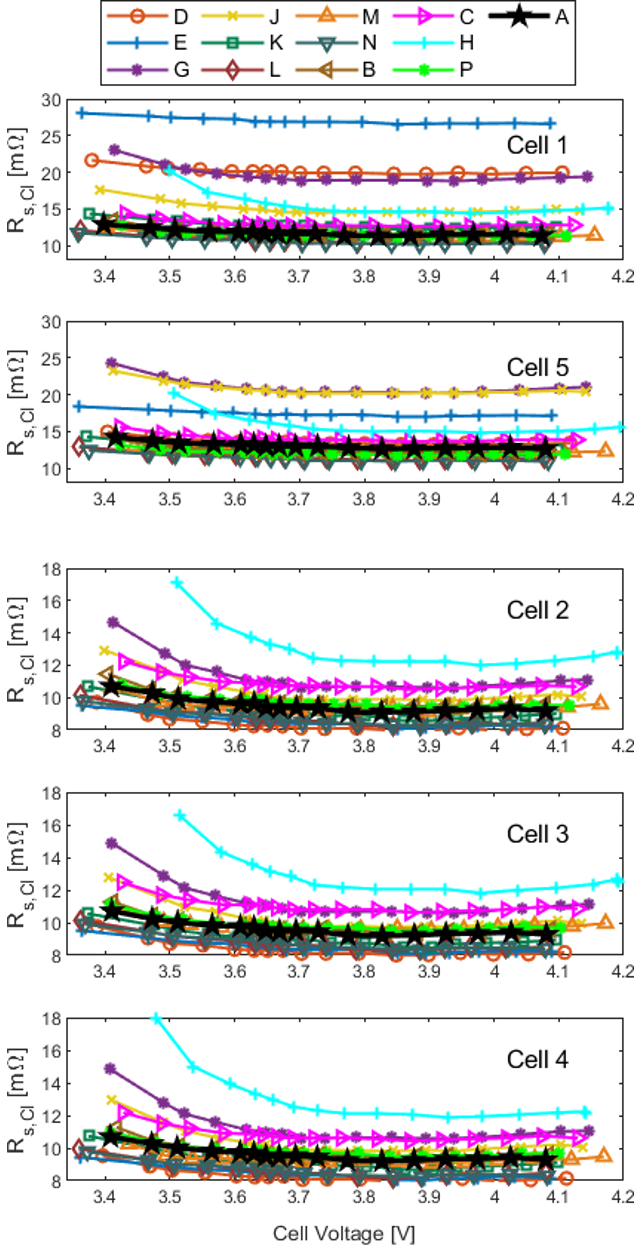}    
\caption{Values of $R_s$ for each of the five cells in a battery pack. Values are taken from charge interrupt test for all thirteen pack analyzed with the fresh pack denoted by A. Cells 1 and 5 are on the exterior of the pack while cells 2, 3, and 4 are on the interior. } 
\label{fig:RS_all}
\end{center}
\end{figure}
Furthermore, the $R_s$ values are relatively constant at mid and high values of the cell terminal voltage. While there is a gradual increase at low voltages, typically around 3.6V or lower, this lower voltage range makes up a significantly smaller portion of the operating range due to the rapid voltage drop near low SOC, as shown in Fig. \ref{fig:Cap_measure}. As such, single charge interrupts after charging to a voltage threshold of approximately 3.6V can provide a rapid approximation for the $R_s$ parameter of a cell.

In Fig. \ref{fig:RS_all}, the $R_{s,CI}$ values from the charge interrupt tests are provided for all five cells in the packs analyzed. The trend of increase $R_{s,CI}$ at lower is observed for all packs, including the fresh pack denoted as A. Further, the increase of $R_{s,CI}$ in cells 1 and 5 show a significant departure from the resistance values measured in cells 2, 3, and 4. However, referring back to the pack schematic provided in Fig. \ref{fig:Packs}, it is assumed that the increase in resistance is not entirely due to internal resistance of cells 1 and 5. Rather, the external resistance of the tabs denoted by $B+$ and $B-$ in Fig. \ref{fig:Packs} may be responsible for a significant portion of the resistance increase. The interior cells 2, 3, and 4 are isolated from this issue as the tabs used to measure the terminal voltage are not current carrying. Further experiments are required to isolate the true internal resistance of cells 1 and 5 from the external tab resistance. All of the following discussion will focus on the results from cells 2, 3, and 4.

From Fig.\ref{fig:HPPCvCI} and \ref{fig:RS_all}, pack H has a noticeably higher resistance than most packs, followed by packs C and G. The other aged packs are more densely populated around the fresh pack A. As the objective of this study is to analyse the ability of $R_s$ to provide an indication of the capacity, a defining metric of SOH, these differences in $R_s$ are next presented with respect to the measured capacity of each pack.

\subsection{Capacity and Series Resistance}
\begin{figure}[t]
\begin{center}
\includegraphics[width=8.4cm]{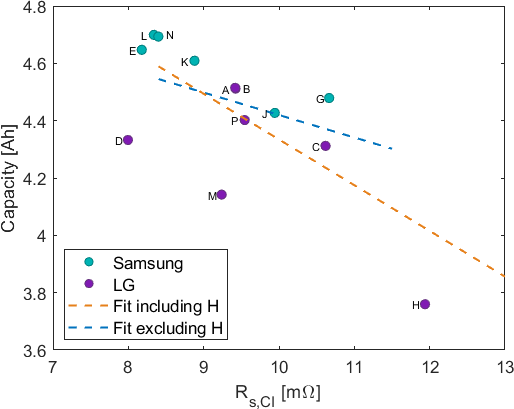}    
\caption{Capacity and resistance for each pack for cell 4. Capacity was measured across a voltage range of 4.1V to 2.95V for C-rate of C/20 and resistance was measured through the use of charge interrupts.} 
\label{fig:CapVRes}
\end{center}
\end{figure}
As outlined in section \ref{sec:test_procedure}, the capacity of the battery packs are calculated between the voltage limits of 2.95V and 4.1V during the the C/20 discharge in TS5, which takes upwards of 20 hours for each pack. In Fig. \ref{fig:CapVRes}, the capacity of cell 4 is shown with respect to the corresponding ohmic resistance $R_{s,CI}$. The $R_{s,CI}$ value used from each pack was the sample point closest to a terminal voltage of 4V. While there is a limited sample size provided for this comparison, the overall relation between capacity and $R_{s,CI}$ can initially be approximated by a linear fit. As mentioned previously, pack H has the highest measured resistance, and this corresponds to the lowest measured capacity for all of the cells. Similarly, pack E, L and N have some of the lowest resistances and highest measured capacities, even higher than that of the fresh pack. However, pack M displays the second lowest capacity while having a resistance almost identical to that of fresh pack A, limiting somewhat the linear relationship between capacity and resistance, along with pack D. 

The proposed linear relationship between $R_s$ and capacity can potentially be explained from a lithium inventory loss perspective as detailed by \cite{Prasad2013}. The solid electrolyte interphase (SEI) layer is an electrically insulating film on the negative graphite electrode in a lithium-ion cell, protecting the electrode from the solvents in the electrolyte. While this layer is largely permeable to lithium ions, the SEI layer will slowly consume lithium as the cell cycles and is considered a major factor in capacity loss for lithium ion batteries. As the SEI layer grows and consumes more lithium, the SEI thickness increases, increasing the internal resistance of the cell and causing more power to be lost (\cite{Plett2015}). 

A linear fit including all of the packs obtains an $R^2$ value of 0.50, which means 50\% of the variance in the data can be explained by the linear fit. While this is a relatively low metric for the goodness of fit, it should be noted that none of the aged pack meet the 80\% capacity metric typically considered for classifying automotive battery packs as below the range of the warranty. Further, if we eliminate pack H from the fit, the general slope of the line is maintained, as shown in Fig. \ref{fig:CapVRes}, but the $R^2$ value drops to 0.26 due to the limited range of pack capacity.  This suggest further analysis with a wide range of aged cells are required to experimentally substantiate the capacity-resistance relationship.

\section{Conclusions}

This paper introduces the methodology for a rapid measurement of ohmic resistance, $R_s$, in a lithium-ion battery and discusses the potential efficacy of using $R_s$ as a metric to estimate the remaining capacity of a cell. From the thirteen battery packs analysed, $R_{s,CI}$ provides an approximately linear indication of capacitance without the necessity evaluating the OVC curve with respect to SOC as shown in Fig. \ref{fig:CapVRes}. 
Further, $R_{s,CI}$ remained constant for mid to high range terminal voltage values, eliminating the need for an exact knowledge of SOC when utilizing $R_{s,CI}$ as a capacity metric. 

Further experimentation with more cells, and in particular more heavily aged cells, is needed to provide a statistical analysis of the $R_{s,CI}$ and capacity correlation of discarded cells.

\begin{ack}
The authors would like to acknowledge the support of Paul Hernley from Battery Solutions, LLC who provided the used battery packs from their recycling facility. 
\end{ack}

\bibliography{main}             

\end{document}